\documentclass[useAMS,usenatbib,usegraphicx]{mn2e}
\onecolumn{}

\title[Least-squared modeling with orbital stability constraints on planets in HU
 Aquarii]{New Light-Travel Time Models and Orbital Stability Study of the Proposed
 Planetary System HU Aquarii}

\author[T. C. Hinse et al.]
       {T.\ C.\ Hinse\,$^{1,2}$,
       J.\ W.\ Lee\,$^{1}$,
       K.\ Go{\'z}dziewski\,$^{3}$,
       N.\ Haghighipour\,$^{4}$
       \newauthor
       C.-U.\ Lee\,$^{1}$, E.\ M.\ Scullion\,$^{5}$
       \\
       $^{1}$\,Korea Astronomy and Space Science Institute, 776 Daedukdae-ro,  
       Yuseong-gu, 305-348, Daejeon, Republic of Korea. (tchinse@gmail.com)
       \\
       $^{2}$\,Armagh Observatory, College Hill, BT61 9DG, Armagh, UK.
       \\
       $^{3}$\,Nicolaus Copernicus University, Torun Centre for Astronomy, PL-87-100 Torun, Poland.
       \\
       $^{4}$\,Institute for Astronomy \& NASA Astrobiology Institute, University of Hawaii, 
       96822 HI, USA.
       \\
       $^{5}$\,Institute of Theoretical Astrophysics, University of Oslo, 0371 Oslo, Norway.}

\usepackage[showcites]{refcheck}
\usepackage{lastpage}

\begin{document}
\maketitle

\begin{abstract}
In this work we propose a new orbital architecture for the two proposed circumbinary planets 
around the polar eclipsing binary HU Aquarii. We base the new two-planet, light-travel time 
model on the result of a Monte Carlo simulation driving a least-squares Levenberg-Marquardt 
minimisation algorithm on the observed eclipse egress times. Our best-fitting model with $\chi_{r}^2=1.43$ 
resulted in high final eccentricities for the two companions leading to an unstable orbital 
configuration. From a large ensemble of initial guesses we examined the distribution of final 
eccentricities and semi-major axes for different $\chi_{r}^2$ parameter intervals and 
encountered qualitatively a second population of best-fitting parameters. The main characteristic 
of this population is described by low-eccentric orbits favouring long-term orbital stability 
of the system. We present our best-fitting model candidate for the proposed two-planet system and 
demonstrate orbital stability over one million years using numerical integrations.
\end{abstract}


\section{INTRODUCTION}

In the past two decades, astrophysical 
timing measurements has been used to infer the existence of multiple low-mass 
planetary objects.  \cite{WolszczanFrail1992} announced the first detection of a 
planetary system around the pulsar PSR1257+12. A two-planet system orbiting the 
short-period subdwarf B of the eclipsing binary HW Virginis was first presented in the 
works by \cite{Lee2009}, and recently \cite{Beuermann2010}, \cite{Potter2011},
and \cite{Doyle2011} announced the existence of two circumbinary planets in possible mean 
motion resonances around NN Serpentis, two giant planets orbiting the eclipsing binary
UZ Fornacis, and a single circumbinary planet around the stars of the binary system Kepler 16, respectively.

The studies by \cite{Beuermann2010} and \cite{Potter2011}
inferred the presence of additional massive objects by 
explaining the observed timing anomalies with the light-travel time 
(LTT, hereafter) effect. In the ideal case, the stellar components of an eclipsing 
binary system are orbiting their common centre of mass with a constant period. 
Timing irregularities can occur if the binary system is accompanied by an 
additional massive object. In this case, the binary centre of mass is orbiting the 
system's centre of mass. At some times the binary will be closer to the observer 
while at other times, it will be farther away, giving rise to the LTT effect on 
measured eclipse egress times.

In a recent study, the measurements of eclipse egress times of the eclipsing 
polar binary HU Aquarii (HU Aqr, hereafter) was used to infer the presence 
of a circumbinary planet around this system \citep{Qian2011}.

These authors modeled the 
complete timing data set by adding the LTT effects from two circumbinary 
planets and found that the pericentre of the outer planetary companion is 
inside the orbit of the inner planet. This orbital architecture implies a 
crossing orbit configuration which points to strong mutual perturbations
between the two planets. 
\cite{Horner2011} subsequently carried out a detailed stability 
analysis of these bodies and found that almost all their initial conditions lead 
to unstable orbits on short-time scales. These authors concluded that the HU Aqr 
planetary system has most likely a different orbital architecture than proposed 
by \cite{Qian2011}.

Three possibilities exist to explain the instability of the proposed planetary system. 
Either i) the LTT parameter space was not explored thoroughly while
orbital stability constrains were imposed [omitted in \cite{Qian2011}], or, 
ii) the applied LTT model was not complete and was unable to explain the timing data set
properly, or iii) the data set is not yet large 
enough to draw reliable conclusions on the existence of
additional low-mass companions.

In this work we aim to find a more plausible orbital architecture that best 
describes the timing data while also being conform with orbital stability 
requirements. We regard the stability condition as an observable that places 
additional constraints on the fitting process \citep{Gozdziewski2005}. We carried 
out a large-scale Monte Carlo, least-squared parameter survey by fitting 
various LTT models to the complete data set. In parallel to the fitting process, 
we performed a stability analysis of various LTT model orbits which best describe the 
data set. By parameterising planet's Hill radii, we imposed orbital stability 
constraints to our parameter survey in order to explore orbital architectures that 
result in stable orbits. We extended the models by \citet{Qian2011} to systems that 
also allow the inner planet to attain an eccentric orbit. Our results suggest that 
two-planet LTT models in general tend to produce orbits with higher eccentricities 
which are likely unstable. 

The outline of this paper is as follows. 
In section 2 we present details of the adopted LTT model and provide a brief 
description of the least-squared minimisation algorithm. In section 3 we introduce 
our orbital stability constrains and in section 4 we describe results of numerical experiments. 
Finally, we give a summary and discussion in section 5.

\section{Analytic LTT Model, Least-squared fitting with orbital stability constraints}
At the basis of our analysis is the combined eclipse egress timing data set
(along with $1\sigma$ error bars) as compiled from the works of \citet{Schwope2001},
\citet{Schwarz2009}, and \citet{Qian2011}. A total of 113 timing measurements were
obtained for HU Aqr which span the time between April 1993 (BJD 2449102.9) and 
May 2010 (BJD 2455335.3), equivalent to a period of approximately
17 years. All timing measurements have been stated in barycentric julian date (BJD) as
described in \cite{Qian2011}, and thus have been transformed to the solar
system barycentre. In the absence of a mechanism that causes timing variations, the 
linear ephemeris is given by
\begin{equation}
BJD = T_{0} + P_{0}E,
\end{equation}
\noindent
where $E$ denotes the ephemeris cycle number, $T_{0}$ is the reference epoch, and 
$P_{0}$ measures the eclipsing period of HU Aqr. The long observing baseline renders 
the eclipsing period to be known with high precision. In this work, we chose 
to place the reference epoch close to the middle of the observing baseline to 
avoid parameter correlations between $T_{0}$ and $P_{0}$ during the fitting process.
In the following, we outline our least-squared fitting technique, and develop 
the LTT model that describes the timing measurements by two independent planets, while 
imposing orbital stability constrains.

\subsection{Analytic LTT Model}
We implemented the LTT model as described in \cite{Irwin1952} in {\sc IDL} and the resulting code is available upon request. In this model, the stars of the binary are assumed to represent 
one single object with a total mass equal to the sum of the masses of the two stars. The assumed 
single object is then placed at the original binary barycentre and in an orbit around 
the system's total centre of mass. In the following, we denote the latter as the LTT orbit. The 
reference system of the model originates from the centre of mass of all involved 
bodies with the $z$-axis pointing towards the solar system's barycentre. The quantity 
$O-C$ for HU Aqr was then calculated using
\begin{equation}
O-C = BJD_{obs} - BJD = \sum_{k=1,2} \tau_{k}(a_{b}\sin       
I_{k},e_{b},\omega_{b},T_{b},P_{b}),
\end{equation}
\noindent
where $BJD_{obs}$ denotes the timing measurement series and, $\tau_1, \tau_{2}$
are given by
\begin{equation}
\tau_{k} = \frac{a_{b}\sin I_{k}}{c}\Big[ \frac{1-e_{b}^2}{1+
e_{b}\cos{f_{b}}}\sin(f_{b}+\omega_{b}) + e_{b} \sin\omega_{b}\Big ].
\label{lightequation}
\end{equation}
\noindent
In Eq.~\ref{lightequation}, the quantity $c$ is the speed of light and $f_b$ is the true longitude
of the binary's centre of mass in the LTT orbit. The LTT orbital parameters are the
projected semi-major axis of the binary barycentre about the total 
centre of mass $a_{b}\sin I_k$, the binary eccentricity $e_b$, 
argument of pericentre $\omega_b$, time of
pericentre passage $T_b$, and the period of the LTT orbit $P_b$. The argument of
pericentre is an inertial angle between the line of apse and a line that is perpendicular to the line connecting the
solar system barycentre and the barycentre of the system (see fig. 1 in Irwin (1952)). We added a subscript to the
inclination as it appears customary and is frequently used in the literature. However, both
objects are orbiting in the same LTT plane. Therefore, they (the merged binary components and the unseen companion) 
share the same inclination.
In general, the line-of-sight inclination of the LTT orbit with respect to the plane 
of the sky $(I_{k})$ is not known from the timing measurements. This is true except for 
cases where the binary companion reveals itself by eclipsing the binary components. 
In this case, the orbits are co-planar (e.g., the circumbinary planetary system of 
Kepler 16) and only minimum separation and minimum mass of the companion are obtained.

To infer the instantaneous angular position of the binary's
centre of mass requires the computation of the eccentric anomaly $E'$ from the mean
anomaly $M$ \citep{MurrayDermott2001}. Adopting the ephemeris cycle number $(E)$ as the independent (time)
variable, the mean anomaly is given by $M = n(BJD - T)$ where $n=2\pi/P$ is the mean-motion
of the combined binary in its LTT orbit. The eccentric anomaly is obtained from the orbital eccentricity and
mean anomaly by solving Kepler's equation. For the purpose of this computation, 
we implemented the Newton-Raphson algorithm to iterate towards a solution for $E'$ 
with an initial starting guess of $E'=M$. The true anomaly is then computed using
\begin{equation}
\tan\frac{f_b}{2} = \sqrt{\frac{1+e_b}{1-e_b}}\tan\frac{E'}{2}.
\end{equation}
\noindent
We tested the correct implementation of the above model by reproducing the light-time 
curves for different values of $\omega_{b}$ as presented in \citet{Irwin1959}.

\subsection{Least-squared Fitting with {\sc MPFIT}}

For the non-linear, least-squared fitting process, we used the 
Levenberg-Marquardt (LM, hereafter) minimisation algorithm as implemented in the 
{\sc IDL} routine {\sc MPFIT}\footnote{http://purl.com/net/mpfit} \citep{Markwardt2009}. 
This algorithm has found widespread applications within astronomy including timing-modeling
\citep{Barlow2011}. In this routine, partial derivatives for computing the gradient field in a
parameter space are calculated from numerical differentiations of model 
functions. The routine also allows for imposing upper and lower boundary 
conditions for each parameter of the model. 
Parameters can be held fixed or flagged as variable as well. In short, 
{\sc MPFIT} attempts to minimise the sum of weighted residuals squared ($\chi^2$) 
between the model and the observed timing measurements. The reduced $\chi^2$ is 
introduced as $\chi_{r}^{2}$ and is given by $\chi^2$ divided by the number of degrees 
of freedom. The algorithm attempts to iterate towards a fit until the difference between 
two consecutive $\chi_{r}^2$ are smaller than some user-defined threshold.

\subsection{The LTT Mass-Function of the Unseen Companion}

In essence, applying the LTT model to an anomalous timing data set provides information
on the binaries barycentre motion and in principle no information is obtained about the
unseen (too dim) companion. However, in the barycentric motion, the orbital period and 
eccentricity (in addition to the inclination of the LTT orbit) are shared between the 
close binary system and the unseen companion. From Kepler's law of motion and 
similar to the case of a single-line spectroscopic binary, the minimum mass of 
the third companion within the context of LTT can be expressed as

\begin{equation}
m_{k}\sin{I_{k}} = \frac{4\pi^2}{G P^2} K c a^2.
\end{equation}
\noindent
In this equation, $m_k$ (where  $k=1,2$) is the mass of the (probable) unseen companion, $G$ is 
the gravitational constant, $K = a_{b}\sin I_{k}/c$ is the semi-amplitude of the observed
light-travel time curve, and $a$ is the minimum semi-major axis of the unseen companion 
\emph{relative to the binary barycentre}. Without additional information about the unseen 
companion it would not be possible to determine its semi-major axis. However, from LTT 
considerations, one can write the mass-function of the companion(s) as

\begin{equation}
f(m_{k}) = \frac{4\pi^2 (a_{b}\sin I_{k})^3}{G P^2} = \frac{4\pi^2 (K c)^3}{G P^{2}} = 
\frac{m_{k}^3 \sin^3 I_{k}}{(m_{b} + m_{k})}.
\end{equation}
This mass-function can be obtained by fitting for $K$ and $P$ 
considering $m_{b}$ as the (combined) mass of 
the two stars. We note that the equation above is transcendental in $m_k$. Assuming that $m_b$ is known, the minimum mass, $m_{k}\sin I_{k}$, of the unseen companion can be determined via iteration. The mass of the binary stars ($m_{b}$) have been reported by \cite{Schwope2011}. The primary white dwarf component was found to be $0.80~(\pm0.04)~M_{\odot}$ while the secondary has a mass of $0.18~(\pm0.06)~M_{\odot}$ 
at an inclination (determined from eclipses) of $87^{+0.8}_{-0.5}$ degrees. Here we use 
$m_b = 0.98~M_{\odot}$. 

It is possible to show that the above transcendental equation has
only one unique solution. In this work, we use the Newton-Raphson algorithm with
$f(m_k)$ as the initial start guess. Finally, it is important to note the the 
LTT-determined semi-major axis and mass of the unseen circumbinary component are only calculated as projected quantities.


%
%

\section{LEAST-SQUAREd FIT WITH ORBITAL STABILITY CONSTRAINTS}

In a similar spirit as outlined in \cite{Gozdziewski2003,Gozdziewski2005}, the dynamical 
stability of a multi-body system is treated as an ``observable''. By ``observable'' we mean 
that a promising best-fitting model to the timing data should result in a stable orbit configuration as 
a necessary condition to substantiate its physical existence. 

To ensure the stability of the system, after each 
successful convergence towards a solution from the LM minimisation procedure, we tested 
the resulting fitted parameters for the following stability conditions:

\begin{itemize}
\item[I.] constraint: $a_{1} < a_{2}$ 
\item[II.] constraint: $q_{2} + n R_{H}^{(2)} > Q_{1} + n R_{H}^{(1)}$, with 
$n=1,2,\ldots$
\end{itemize}
\noindent
where $q_{2}=a_{2}(1-e_{2})$ and $Q_{1}=a_{1}(1+e_{1})$ are the pericentre and apocentre
for the outer and inner planet, respectively. The modified Hill radius for each planet is given by
\begin{equation} 
R_{H}^{(k)} = a_{k}^{*}\Big(\frac{m_k}{m_{b}+m_{k}}\Big)^{1/3}
\end{equation}
\noindent
and a multiple $n$ of this quantity is used to parametrise the stability condition. 
It is important to note that $a_{k}^{*}$ is the semi-major axis of the two proposed 
planetary companions \emph{relative to the binary's centre of mass} (still treated as a 
single object with mass $m_{b}$). At this point we note that we use a different 
definition of the Hill radius than used in \cite{Horner2011}. Our definition (although somewhat similar) results in a larger (around 40\%) radius imposing a more stringent initial spacing condition between the two planets.

From the two above-mentioned stability constraints,
the first condition simply demands a hierarchical system with the size of the inner 
planet's orbit always being smaller than that of the outer body. For eccentric orbits,
we use the ``spacing parameter'' $n$ to quantify a stability condition. The second
condition essentially demands that the outer planet's pericentre distance to be always 
larger than the inner planet's apocentre. This condition is clearly violated for the 
case of the planetary orbits in HU Aqr as proposed by \cite{Qian2011}. In an attempt to 
avoid close encounters, the condition is augmented by demanding that the spacing between 
the planets to be a multiple of their respective Hill's radii. For small values of $n$, mutual 
gravitational interactions are most likely to disrupt the system. Intuitively, for large 
values of $n$, the two orbits are more apart from each other. Hence imposing a larger 
spacing condition would constrain the parameter space considerably while possibly allowing 
for stable orbital architectures. To obtain stable orbits, 
after each iteration of the least-squared 
fitting procedure, we determine the minimum mass of the two companions. Combined with 
the orbital period, we then calculate the minimum semi-major axis $a\sin I_{1,2}$ of the two 
planets from Kepler's third law. Finally, we evaluate the two stated stability conditions 
for certain orbits obtained from the fitting process. If a given set of fitted parameters 
passes the stability conditions, the resulting orbits are numerically integrated to 
directly infer their long-term stability. In this work we have used $n=1$ as will be 
justified later.

\section{METHODS, NUMERICAL EXPERIMENTS, AND RESULTS}

\subsection{Lomb-Scargle Frequency Analysis}

\begin{figure*}
\includegraphics[]{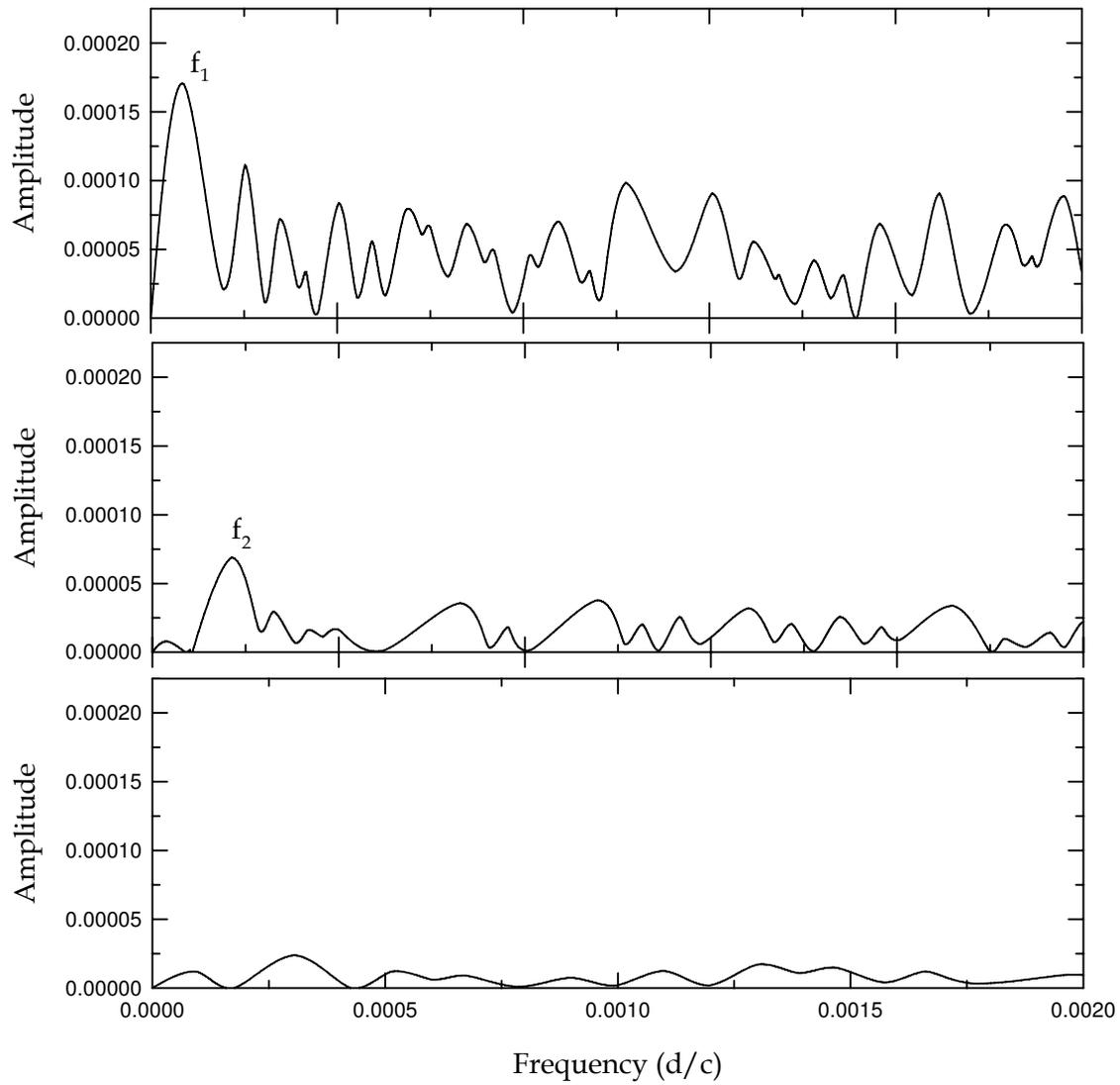}
\caption{Lomb-Scargle periodogram of the $O-C$ eclipse egress timing measurements using the complete data 
set from the literature. Two distinct
frequencies are found with $f_{1} = 0.000176$ cycles/day (15.6 years) and $f_{2} = 0.000425$ cycles/day 
(6.5 years). The amplitude of $1/f_1$ is larger because it corresponds to the period of the wide orbit 
companion.}
\label{periodogram}
\end{figure*}

In this section, we describe our numerical experiments, the applied methods, and final outcome. As a 
first step, we applied a periodogram analysis to our data set 
using the \texttt{PERIOD04} algorithm. This analysis enables us 
to extract the frequency content in an unevenly spaced data set \citep{LenzBreger2005} 
and also allows for an estimation of any periodicity that may exist in the data 
which can be used to provide an initial guess for the 
least-squared fitting procedure. At this stage, we are not aiming at accuracy.
From the complete data set, we constructed the $O-C$ diagram of HU Aqr at 
the measured times of eclipse egress, and used the period analysis algorithm. 
The resulting Lomb-Scargle powerspectrums are shown in Fig.~\ref{periodogram}. 
Two periods can clearly be seen in the data. 
A long-period variation of approximately 15.6 years, and a shorter 
period of 6.5 years. These values reflect the possible periods of two independent LTT orbits. 
Given the total observation time span of about 
17 years, the shorter period is well-determined. This period has less power which implies a 
smaller orbit relative to the long period one. Later we will demonstrate that the timing 
data cannot be described by a single sinusoidal period alone. Therefore, both periodic signals 
must to some degree be real.

\subsection{Numerical Monte Carlo Experiments and Best-fitting Model}

In principle, searching for and finding a global minimum requires the probing of a large 
number of initial guesses spread across the $\chi_{r}^2$ parameter space. Despite the vastness of this
space, it is possible to limit the search 
by extracting additional information about some of the parameters of the system
from the timing data set. For 
instance, the preceding period analysis can be used to obtain reasonable starting guesses for 
the fitting algorithm. Also, the model parameter $a_{b}\sin I_{1,2}$ can be estimated 
from the largest amplitude in the data set. We find $K \approx 3.47 \times 10^{-4}$ days as a 
maximum which translates into a projected semi-major axis of $a_{b}\sin I_{2} = 0.06$ au for 
the orbit of the outer planet. We carried out 
a ``brute-force'' search for a best-fitting model using a Monte Carlo approach. Initial guesses for the 
parameters of the model were then drawn from a random uniform and/or normal (Gaussian) distribution. 
The projected semi-major axis, argument of pericentre, and orbital eccentricities were chosen 
randomly from a uniform distribution with $a_{b}\sin I_{1,2} \in [0,0.060]$ 
au, $\omega_b \in [-\pi,\pi]$, and $e_b \in [0,0.70]$. We assigned an upper limit on the orbital 
eccentricities to explore different possibilities of orbital architectures while evaluating 
their goodness-of-fit to the data. The times of pericentre passages were also chosen at random 
from a uniform distribution with initial guesses bounded by the earliest and latest epochs of 
observation. Since we have a good idea of the initial reference epoch, eclipsing period, and 
the planet's orbital period, we decided to choose these values from a normal distribution. 
The mean and standard deviation for the reference epoch $T_{0}$ was BJD 2452174.36592 and 0.0001 
days, and for the orbital period $P_0$ was 0.08682040612 days and $10^{-10}$ days, respectively. The mean 
of the LTT period $P_0$ was obtained from the Lomb-Scargle frequency analysis with standard 
deviation of two years for each planetary companion. In order to avoid parameter correlations 
between the reference epoch and the eclipsing period, we chose the reference epoch $(E=0)$ to 
be in the centre of the data set. \cite{Qian2011} chose to place the reference epoch in their 
analysis at the first timing record.

We modeled the complete data set (as obtained from the literature) using a two-Keplerian LTT model. 
In their analysis and by construction, \citet{Qian2011} forced the inner companion to be in a circular 
orbit. While this requirement reduces the number of free parameters, it does not account for the 
possibility of eccentricity excitation of the inner planet by an outside perturber 
\citep{MalmbergDavies2009}. In our study, we allowed the inner planet to also attain some eccentricity. It 
is however, important to point out that in the two-Kepler LTT model, no mutual gravitational 
perturbations are included. In our first approach to find a best-fitting model, we generated 111,844 initial 
guesses. Each LM 
guess was allowed a maximum of 500 iterations prior to termination. We carried out the Monte
Carlo simulations on a supercomputer lasting for about 14 days of non-stop computing. In detail, 
we used the two-sided numerical derivative option available in {\tt MPFIT} algorithm. We also chose 
to assign individual maximum step-sizes to each model parameter. For example, we allowed the 
eccentricity to only change by at most 1\% per iteration. Maximum change in an angular quantity 
was $5^{\circ}$. Similar reasonable values were assigned to other parameters. We believe that 
our initial choices of the value of the parameters renders the search-space of $\chi_{r}^2$ to have a
fine enough grid for each iteration of the fitting algorithm. We also believe that with such a large
number of initial guesses the full $\chi_{r}^2$ space will be sufficiently explored. 
During the fitting process, we held all twelve parameters of the system free, and decreased the 
numerical accuracy 
parameters in {\tt MPFIT} from their default values. We also decreased the parameter step-size with 
which the numerical derivatives were evaluated.

We then carried out a Monte Carlo experiment based on a total of 111844 randomly distributed 
initial guesses. For each converged initial guess with $\chi_{r}^2 \le 10.0$, we recorded the 
resulting $\chi_{r}^2$, the number of iterations, the initial guess parameters, and the final 
converged model parameters as returned by the LM minimisation algorithm. In order to avoid any 
bias effect, we removed converged iterations that reached a lower or upper boundary value. 
That means, once a parameter reached any of the boundaries within its defined range (see above), 
its value was held constant and the parameter was no longer changed  by the minimisation algorithm
for carrying out the remaining of the iterations. Although we have no strict explanation 
for this ``stickiness'' effect, we judge that it has the potential to skew the final distribution 
of model parameters. The idea is that from a mathematical point of view, the minimisation algorithm 
attempts to cross the lower or upper boundary to obtain a better fit based on the steepest descent 
in parameter space. Hence it remains constant at these locations but with no physical meaning. 
As an example, one can consider the orbital eccentricity. Since in the systems of our interest 
(circumbinary planetary systems) eccentricities smaller than zero 
do not correspond to acceptable orbits, iterations that were stuck at the lower boundary of $e=0$ 
were removed. This allows us to avoid introducing a bias towards a preferred direction in parameter 
space that has no reasonable physical significance. In addition, {\tt MPFIT} algorithm does not return any 
formal uncertainties for fitted parameters 
that reach a lower or upper boundary. 
After this cleaning process, we ended up with a data file containing a total of 48485 initial guesses.

\begin{figure*}
\includegraphics[scale=0.5]{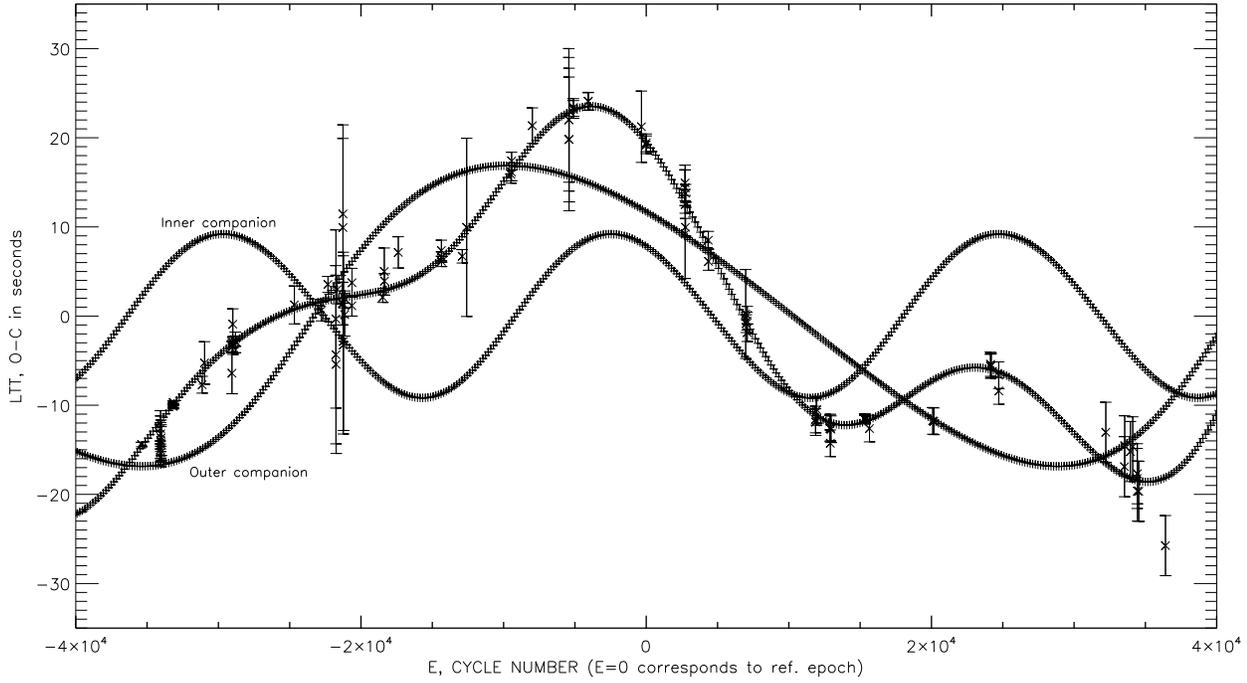}
\caption{Best-fitting two-Kepler LTT model with reduced $\chi_{r}^2=1.43$. The two single  LTT sinusoids 
are superimposed. The inner companion is on a smaller orbit with a smaller amplitude. The smallest and 
largest timing errors are 0.10 and 9.99 seconds, respectively. The group of data at cycle number 34000 
are from \citet{Qian2011}.}
\label{BestFitOmC}
\end{figure*}

Our best fit resulted in a $\chi_{r}^2 = 1.43$ which had an occurrence frequency of fourteen 
times in total. We show the corresponding $O-C$ diagram of this case in Fig.~\ref{BestFitOmC}. 
Fitted model and derived parameters are shown in Table~\ref{fitparamtable} with formal uncertainties 
as obtained from the solutions covariance matrix. We point out the close agreement between the two 
fitted periods (6.5 and 15.5 years, respectively) and the 
periods obtained from the Lomb-Scargle frequency analysis shown in Fig.~\ref{periodogram}. We 
visually inspected the remaining thirteen models and did not find any differences. All final fitted 
parameters agreed with one another. 
However, we noticed a qualitative difference when comparing our best fit with Fig.~2 in 
\citet{Qian2011}. These authors show a better description of the newly added timing data but 
they do not provide a quantitative $\chi_{r}^2$ value for their best fit. When comparing our 
results, we noticed that the inner companion 
in our model can attain eccentric orbits, which might explain the difference between the fits 
presented in this work and in \cite{Qian2011}. In addition, in Fig.~\ref{BestFitOmC}, it appears 
that a single LTT (sinusoidal) curve will provide a poor description of the timing data. 
Neither the individual short (inner planet) nor the long-period (outer planet) LTT model will be able
to provide a satisfactory fit by itself. Only by combining the two LTT models one can obtain
a satisfactory fit.

\subsection{Orbital Integration of Best-fitting Model}

\begin{figure*}
\includegraphics[angle=-90,scale=1.0]{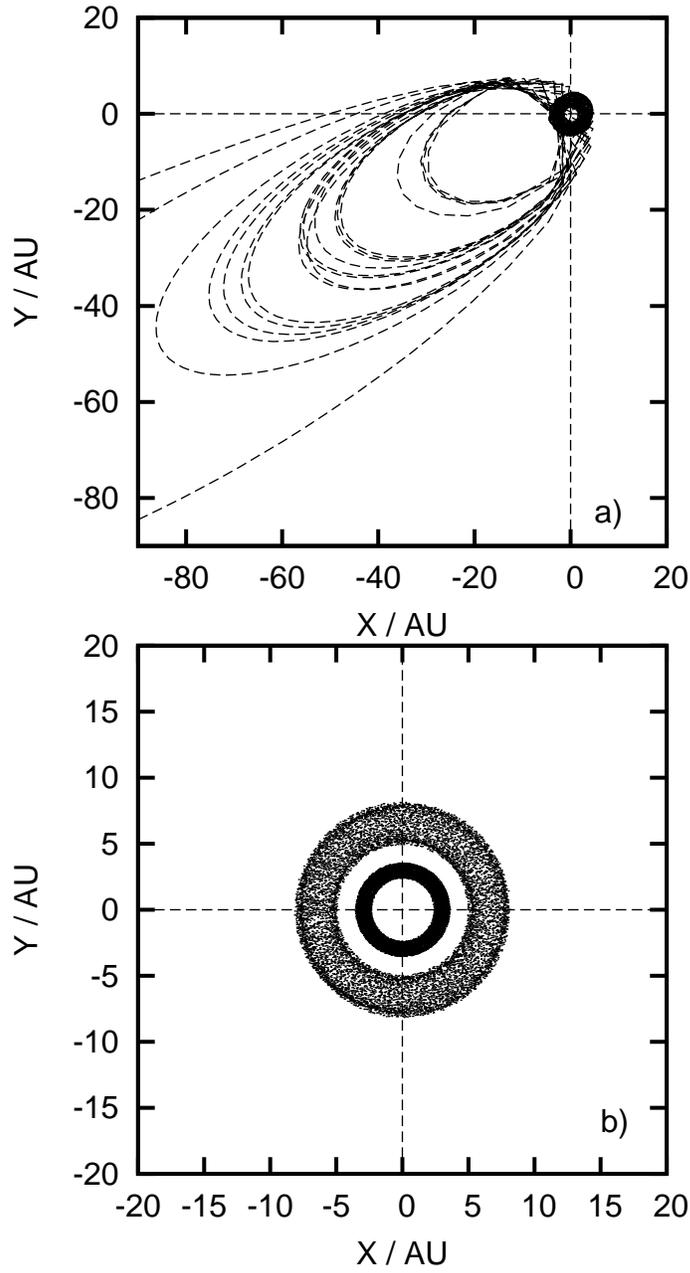}
\caption{Orbital stability of the two companions. \emph{Upper panel}: Unstable system corresponding 
to the nominal case of best-fitting $(\chi_{r}^2 = 1.43)$ model. 
\emph{Lower panel}: Stable system for the optimistic parameters (see text).}
\label{TwoOrbitFigs}
\end{figure*}

Our final orbital parameters for the two planetary companions are relatively close to ones reported 
by \cite{Qian2011} except for a slight increase (decrease) in the inner (outer) companion's orbital 
eccentricity. The outer companion's 
mass and semi-major axis are also larger in our work. We examined the orbital stability of our 
best-fitting model 
(Table \ref{fitparamtable}) using the orbital integration package 
{\tt MERCURY}\footnote{www.arm.ac.uk/$\sim$jec} \citep{MERCURY-1, MERCURY-2}. We integrated the 
three-body equations of motion using the variable-step Bulirsh-Stoer (BS2) algorithm. Initial 
step-size was chosen to be 1\% of a day and the accuracy parameter was set to $10^{-16}$. To be 
consistent with the fitted LTT orbits, we considered the binary pair as 
one single object with a mass equal to the sum of the masses of the two stars. 
We carried out integrations for two scenarios: i) the nominal parameters for the best-fitting model with $(\chi_{r}^2 = 1.43)$, and ii) an optimistic scenario where we considered the lower 
limit in masses and eccentricity and upper limit in the semi-major axis for both companions. 
We integrated both systems for $10^4$ years. The results are shown in Fig.~\ref{TwoOrbitFigs}. This short 
integration time is sufficient to demonstrate an unstable configuration for the first scenario. 
The upper panel of the figure shows the orbits based on the nominal best-fitting parameters and clearly 
shows orbital instability of at least one of the planet companion that resulted in its eventual escape.

\begin{figure*}
\includegraphics[angle=-90,scale=1.0]{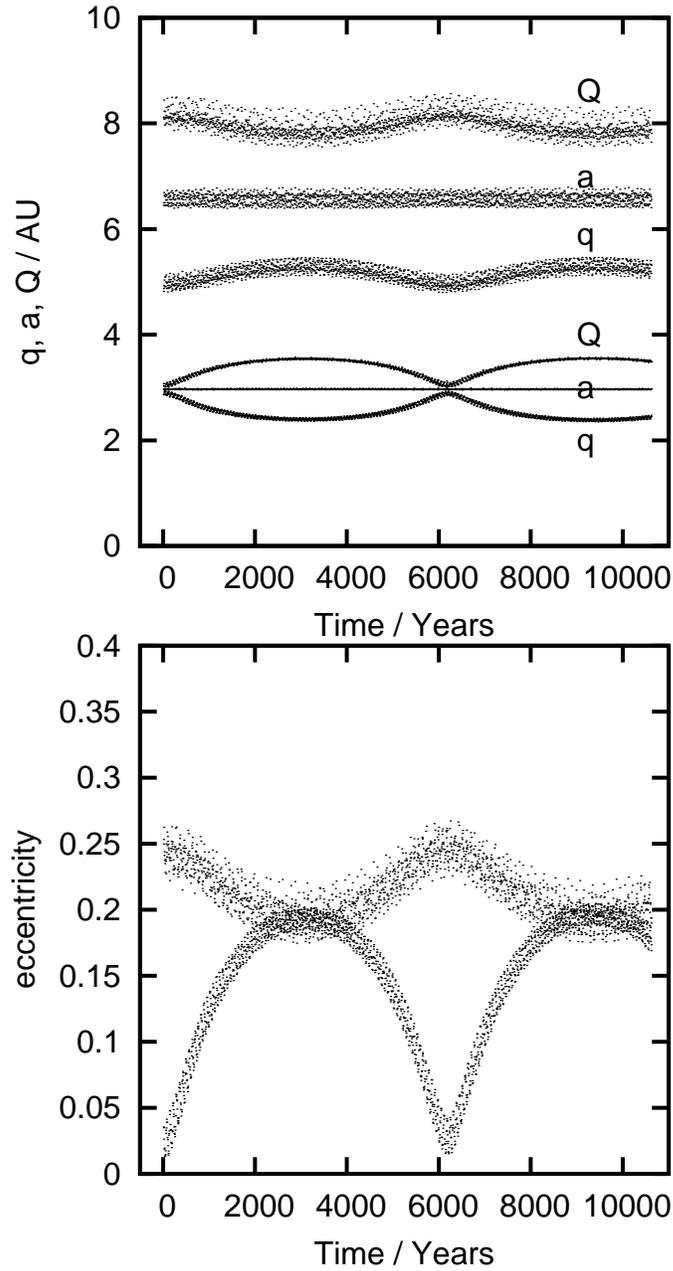}
\caption{Time evolution of orbital elements for the best-case scenario as shown in 
Fig.~\ref{TwoOrbitFigs}b (see text). \emph{Top panel}: Periastron $(q)$, apastron $(Q)$ distance 
and semi-major axis $(a)$. The Hill radius for the inner and outer companion was determined to be 
0.60 and 1.10 au, respectively. \emph{Bottom panel}: eccentricity evolution. The inner companions 
eccentricity varies between circular and moderately eccentric orbits. Although we only show a 10000-year 
time evolution, the two orbits are stable over $10^6$ years.}
\label{TwoOrbitQAEFigs}
\end{figure*}

The optimistic scenario shown in the lower panel of Fig. 3 resulted in a relatively stable orbital 
architecture when compared with the nominal scenario. The corresponding Kepler elements of the orbit
in this case and their time evolutions are shown in Fig.~\ref{TwoOrbitQAEFigs}. 
From this figure we note that both our 
initial stability conditions are satisfied: i) the inner companion has a smaller semi-major axis 
than the outer planet and, ii) the quantity 
$q_{2} + R_{H}^{(2)} \approx 5.9~\textnormal{au}$ stay always larger than 
$Q_{1}+R_{H}^{(1)}\approx$  4.2 au. Also
in this scenario, the two companions are well separated from one another. 
The latter justifies our choice of $n=1$ for the stability 
spacing parameter as introduced earlier. 
As shown in the top panel of Fig. 3, the parameters of our nominal best-fitting model resulted in an 
\emph{unstable orbital configuration}. Combined with the results of the optimistic scenario experiment, 
this suggests that stable systems are to be found among less eccentric orbits for the inner and/or 
outer companion.

\subsection{$\chi_{r}^2$ Frequency Distribution}

\begin{figure*}
\includegraphics[]{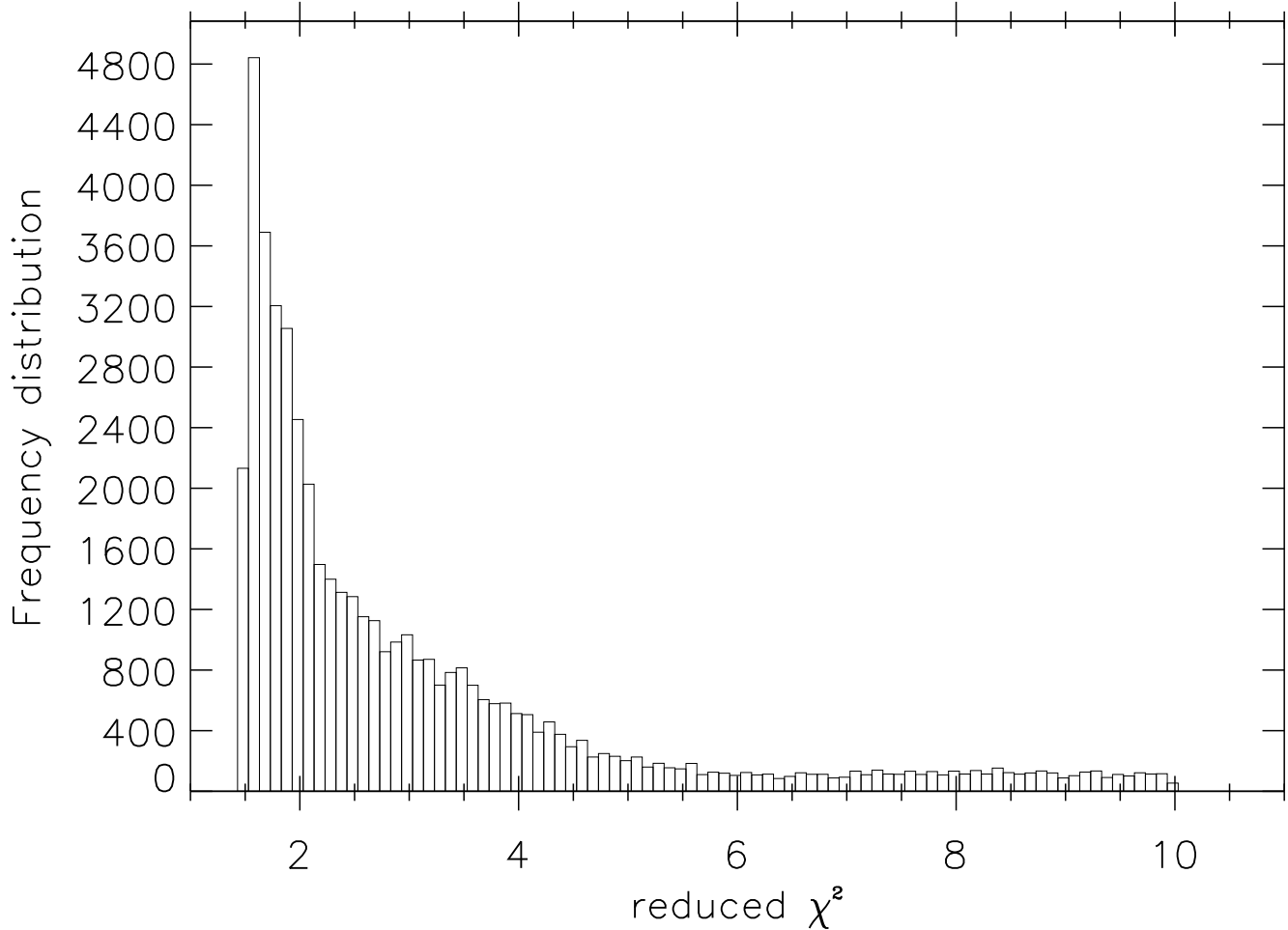}
\caption{Histogram of reduced $\chi_{r}^2$ values based on a 111844 guess Monte 
Carlo experiment. The shown binsize is 0.10. The smallest and largest $\chi_{r}^2$ are 1.43 and 9.99, 
respectively. The bins are centered at mid-bin values which explains the last bin to exceed 
$\chi_{r}^2 = 10.0$. Summing all bins results in 48485 guesses (see text).}
\label{chi2histogram}
\end{figure*}

In order to find  model parameters that result in a stable orbital configuration, we closely examined 
the frequency distribution of the $\chi_{r}^2$ values. Prior to our experiment, we had expected a 
$(1/\chi_{r}^2)$-like histogram 
distribution. In this distribution the majority of final $\chi_{r}^2$ values will belong to the first 
bin with smallest $\chi_{r}^2$. In Fig.~\ref{chi2histogram} we show the histogram with a bin 
size of 0.10 that was obtained from our actual results. As is evident from this figure, increasing the 
bin size would result in the expected 
$1/\chi_{r}^2$ distribution. However, we noticed that unlike what is expected from this type 
of distribution, a non-negligible fraction (4\%) of initial 
guesses produce models with final fits in the interval 
$1.43 \le \chi_{r}^{2} < 1.53$. The histogram's peak value is associated with the interval 
$1.53 \le \chi_{r}^{2} < 1.63$ amounting to 10\% of the total number of guesses. Technically, our 
best-fitting model with $\chi_{r}^2 = 1.43$ represents the global minimum in the $\chi_{r}^2$-parameter space, 
and hence gives the most precise description of the data. Though, the majority of initial guesses 
have a higher frequency at somewhat larger $\chi_{r}^2$ values.

\begin{figure*}
\includegraphics[scale=0.7]{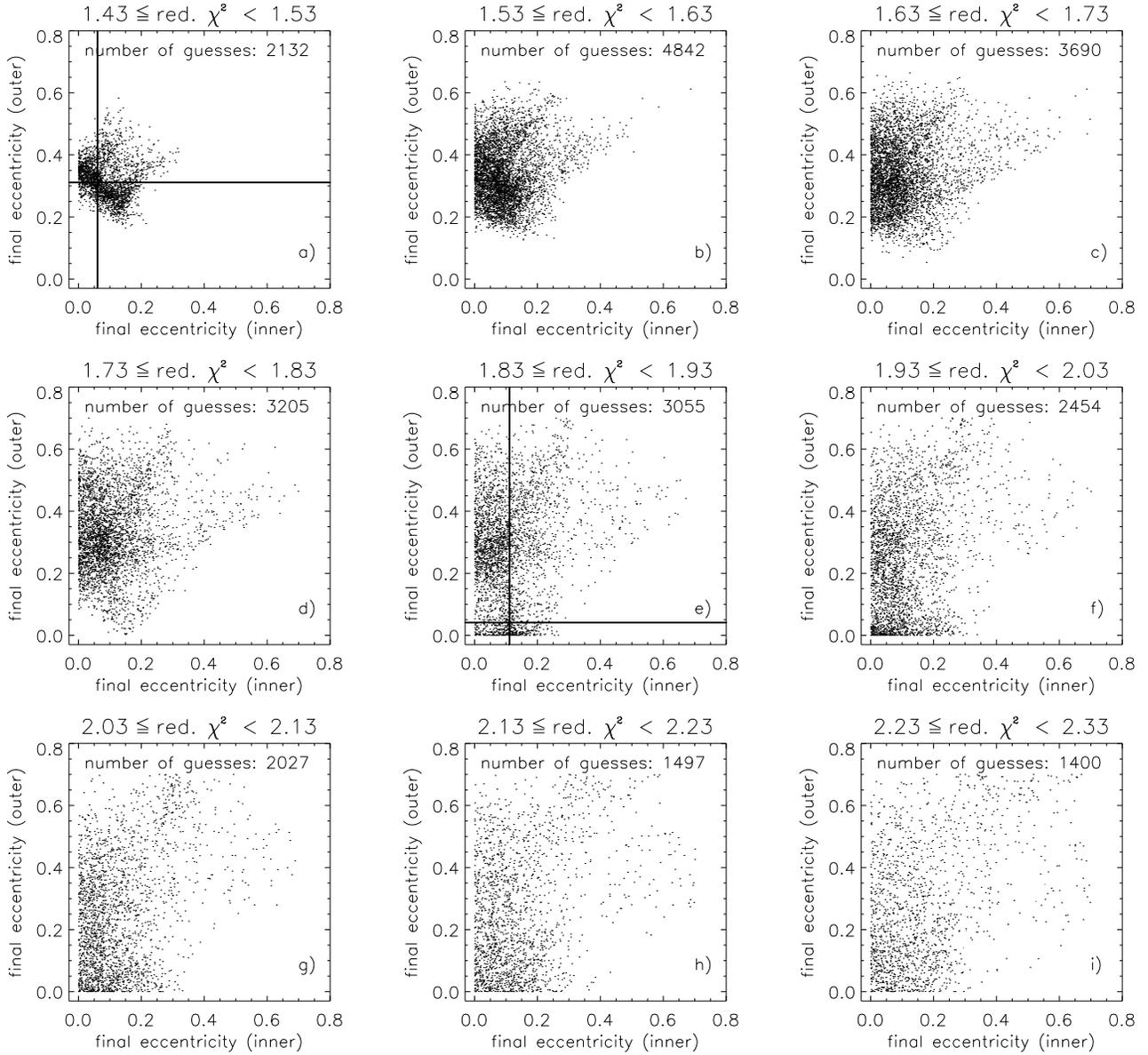}
\caption{Scatter plots of final fitted eccentricities for different $\chi_{r}^2$-bin intervals following 
the histogram distribution shown in Fig.~\ref{chi2histogram}. An apparent bimodal distribution is found 
for the bin $1.83 \le \chi_{r}^2 < 1.93$. We note that all orbits have $0.0 < e_{1,2} < 0.70$ since we 
removed those initial guesses that ``stick'' to the lower and upper boundary of final eccentricity during 
the minimisation/iteration process.}
\label{e-distribution}
\end{figure*}

\begin{figure*}
\includegraphics[scale=0.7]{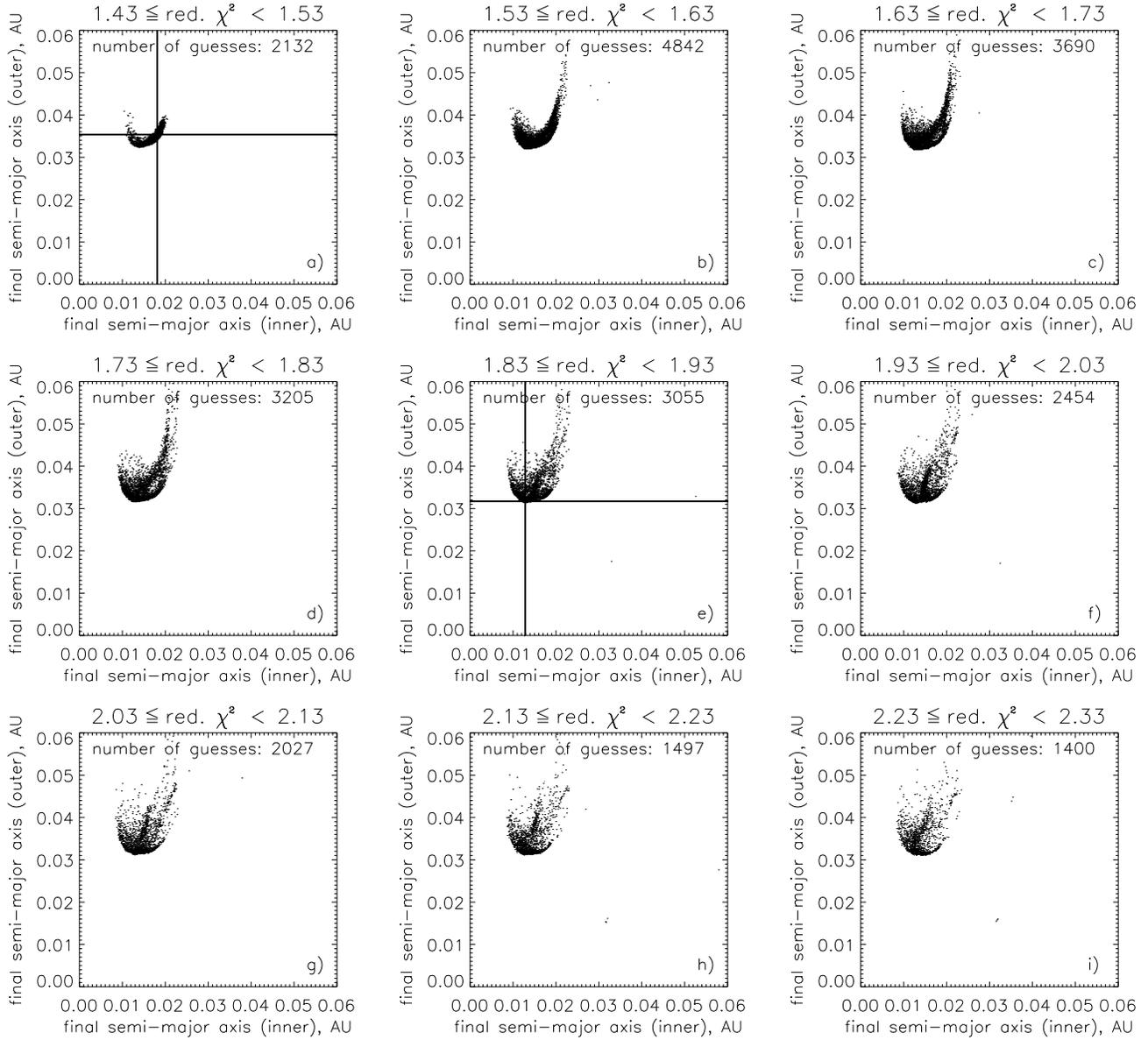}
\caption{Same as Fig.~\ref{e-distribution} but this time for the final semi-major axes of the two proposed 
planetary companions. The semi-major axis is the fitted minimum semi-major axis $a_{b}\sin I_{1,2}$.}
\label{a-distribution}
\end{figure*}

We also examined the distribution of the final fitted eccentricities $(e)$ and minimum semi-major axes 
$(a_{b}\sin I_{1,2})$ for the two companions for all bins shown in the histogram of 
Fig.~\ref{chi2histogram}. In Fig.~\ref{e-distribution} we show the resulted distribution for the 
final eccentricities and in Fig.~\ref{a-distribution} for the final semi-major axes. For the first 
bin interval (Fig.~\ref{e-distribution}a) starting with $\chi_{r}^2 = 1.43$, our simulations indicate 
that the final fitted eccentricity for the outer companion is in general high with a minimum at around 
0.20 and a maximum at 0.50. The final eccentricity for the inner planet ranges from $\approx 0.0$ to 
0.30. The corresponding final fitted semi-major axes appear to be less scattered with final fitted 
values forming an arc-like feature as shown in Fig.~\ref{a-distribution}. In particular, we note a 
lower limit on the projected semi-major axis of the outer planet equal to 0.0312 au. The inner 
companion's semi-major axis appears to be constrained between 0.008 and 0.02 au. 
The cross hair in the upper left panel in Figs.~\ref{e-distribution} and \ref{a-distribution} 
represents the location of the average of all fourteen best-fitting models (with $\chi_{r}^2=1.43$) final minimum 
semi-major axes and eccentricities. Based on our previous orbital stability analysis of the nominal 
best-fitting orbits, we estimate that the final fitted parameters obtained from the first $\chi_{r}^2$ bin, 
will result in unstable orbits because of large orbital eccentricities that cause close approaches 
between the two planets.

\subsection{New System Parameters and Orbital Stability Constraints}

For the first three bin intervals in Fig.~\ref{e-distribution}, the final eccentricities scatter over 
a larger range. The majority of the initial parameter guesses (4842 in total) ended up with 
$\chi_{r}^2$ between 1.53 and 1.63. Our results point to a qualitative emergence of an apparent 
bimodal distribution for the bin interval $1.83 \le \chi_{r}^2 < 1.93$. A density enhancement is 
observed for final eccentricities centered at $(0.07,0.3)$ and a second one at $(0.12,0.01)$. 
Although these final fitting parameters correspond to a poorer goodness-of-fit quality, they allow 
near-circular orbits for the outer companion and hence would satisfy our stability constraints as
outlined earlier.

\begin{figure*}
\includegraphics[scale=0.5]{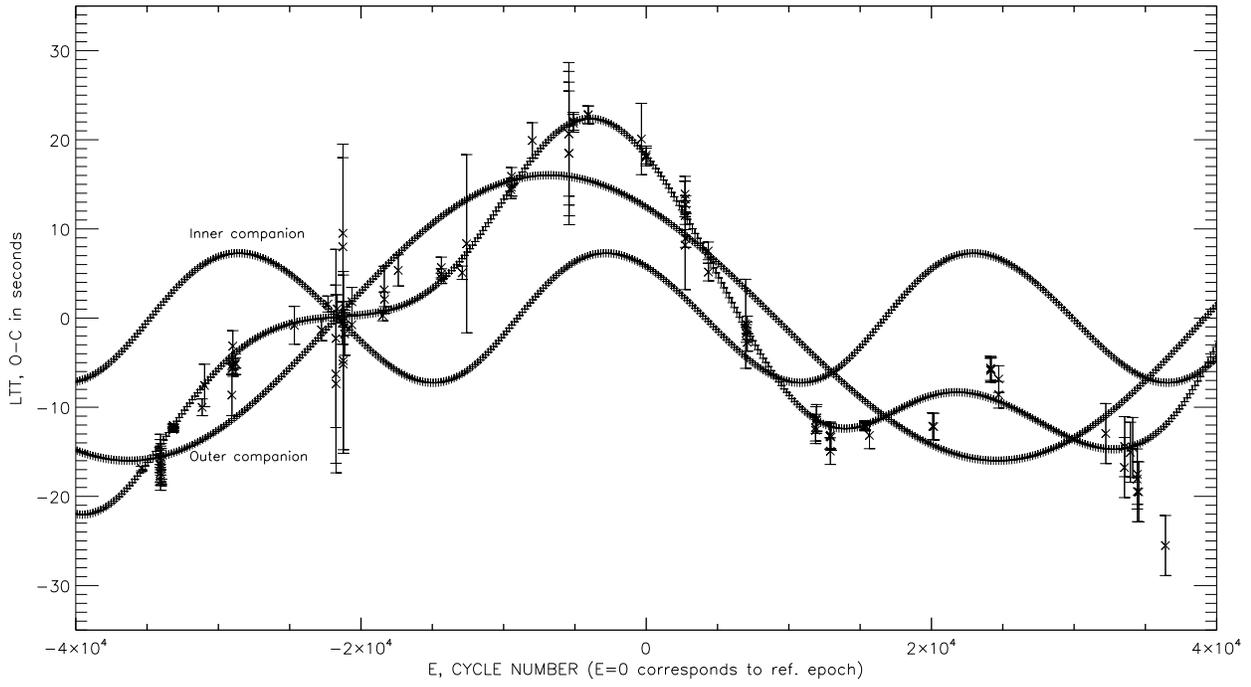}
\caption{Best-fitting two-Kepler LTT model with reduced $\chi_{r}^2=1.89$. The best-fitting model parameters are 
shown in Table~\ref{fitparamtable_2}. The final eccentricities and semi-major axes are indicated by a 
cross hair in Figs.~\ref{e-distribution} and \ref{a-distribution}.}
\label{BestFitOmC_2}
\end{figure*}

An inspection of Fig. 7 indicates that the bin interval $1.93 \le \chi_{r}^2 < 2.03$ also seems to result 
in circular orbits for 
both companions, showing an enhancement of final eccentricities around (0,0). We would like to stress 
that these findings are of qualitative 
judgment from visual inspections. We tested the orbital stability of one of the final parameters with 
$1.83 \le \chi_{r}^2 < 1.93$. The final eccentricity of the two planetary orbits were close to circular. 
We have shown the locations of these orbits with a haircross in Figs.~\ref{e-distribution}e and 
\ref{a-distribution}e for their corresponding semi-major axes. The associated fitted model had 
$\chi_{r}^2=1.89$ and is shown in Fig.~\ref{BestFitOmC_2}. Table~\ref{fitparamtable_2} shows the 
corresponding numerical values of the system's parameters. Comparing these results with those shown 
in Fig.~\ref{BestFitOmC} indicates that the qualitative differences are minute but with dramatic 
consequences for the underlying stability of the proposed circumbinary planetary system. 

\begin{figure*}
\includegraphics[scale=0.8]{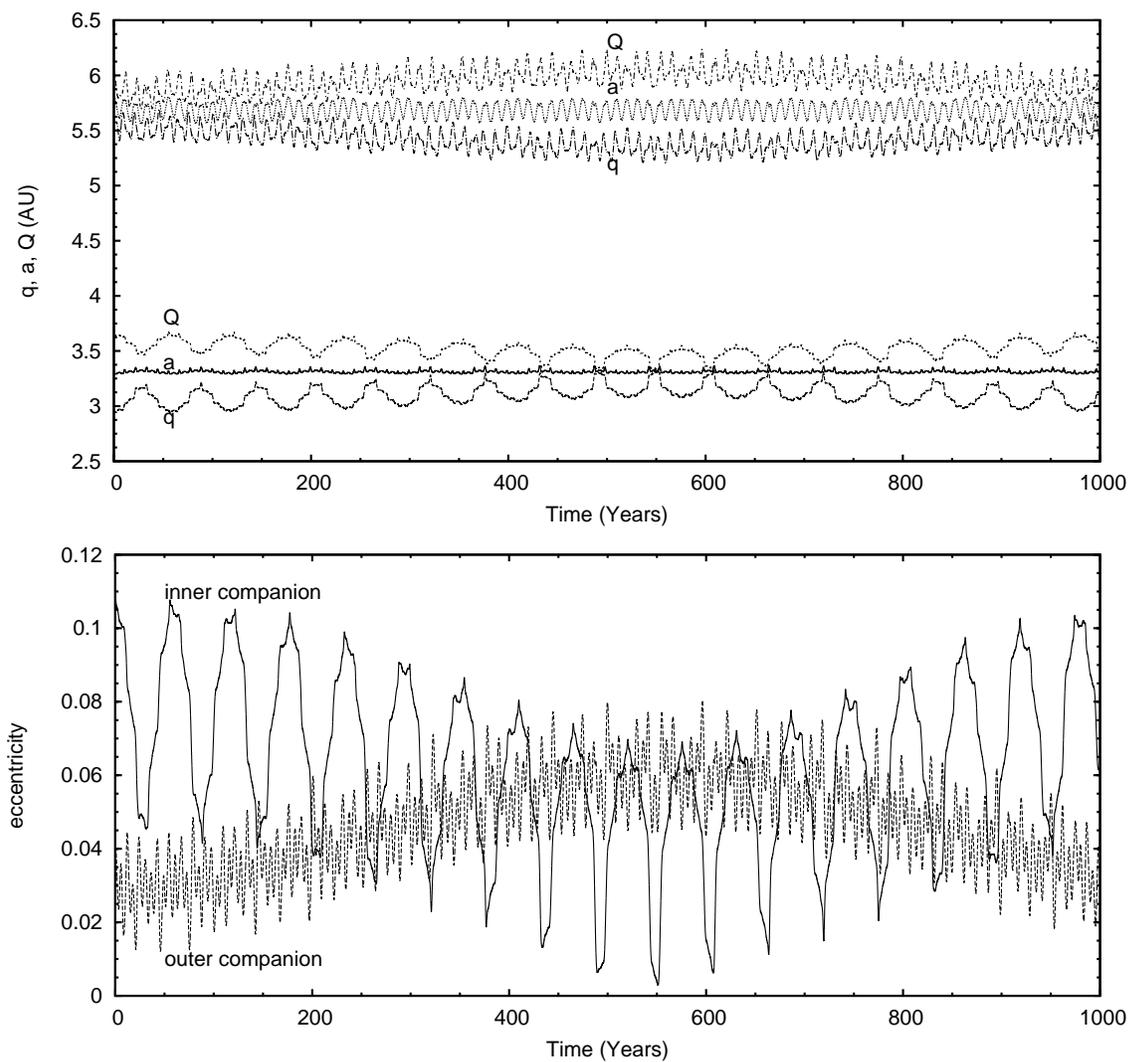}
\caption{Result of a 1000-year orbital integration of the two proposed companions for the best-fitting initial 
condition with $\chi_{r}^2=1.89$. The integration was extended to $10^6$ years and showed bounded/stable motion. 
\emph{Upper panel}: Time evolution of periastron, semi-major axis and 
apastron distance. \emph{Bottom panel}: Time evolution of eccentricity.}
\label{BestFitStability}
\end{figure*}

We also tested the orbital stability of the model with $\chi_{r}^2=1.89$. 
The result of a 1000-year integration is shown in Fig.~\ref{BestFitStability}. 
We show the pericentre, semi-major axis, and apocentre distances in the top panel and the eccentricity 
in the bottom panel. The fitted parameters in this system satisfy the two stability conditions outlined 
before. While the semi-major axis exhibits short-term variations, the variations in the two eccentricities 
are long-term and secular. This is possibly due to exchange of orbital angular momentum between the two 
planetary companions. We also carried out the integration over a $10^6$ year time interval and found 
the orbits 
to be confined and stable with no signs of global instabilities. Particularly interesting is that the 
$\chi_{r}^2=1.89$ fit parameters do not show an orbit-crossing architecture. To substantiate our results, 
we examined several other orbital parameters with goodness-of-fits in the neighbourhood of 
$\chi_{r}^2=1.89$. All of the final parameters resulted in stable orbits over a $10^6$ year integration 
period. Carrying out a detailed dynamical analysis of the orbits shown in Fig.~\ref{BestFitStability} 
is beyond the scope of this work.

\section{SUMMARY AND DISCUSSION}

In a recent study \cite{Qian2011} announced the discovery of a two-planet system around the eclipsing 
polar binary HU Aquarii. In a subsequent paper, the stability of their best-fitting orbital parameters was 
examined by \citet{Horner2011} who showed that the system is highly unstable. The most likely reason for this instability 
is the high eccentricity of the outer companion. In an attempt to find a system with stable orbits, we 
calculated a large number of two-planet LTT models 
using a Monte Carlo approach. This approach provides statistics on the distribution of the final 
goodness-of-fit $\chi_{r}^2$ parameter in combination with constrains on requiring initial orbital 
stability. 

In this work, we analyzed the complete available timing data of HU Aquarii. We made the following assumptions
in our analysis.
First we allowed the inner planet to attain an eccentric orbit in response 
to possible eccentricity excitation by an outside perturber. Compare to the study by \cite{Qian2011},
this assumption resulted in introducing three extra free parameters. We also did not include a secular term 
which would account for timing variations due to mass-/angular momentum transfer and/or magnetic activity 
of the binary component. \citet{Qian2011} argue that the secular term in their work is most likely caused 
by the presence of a third circumbinary massive object. However, the existence of this body in their data 
is far from obvious as the observational time span is much smaller than the orbital period of the 
proposed third circumbinary (planetary) companion. The confirmation of the third planet would require a 
few decades of future photometric follow-up monitoring. Finally, we did not consider mutual gravitational 
interactions between the two planets or gravitational perturbations on the binary stars.

Our best fit model had a goodness-of-fit parameter of $\chi_{r}^2=1.43$ occurring fourteen times out of 111.844 
initial guesses. From studying the histogram distribution of final $\chi_{r}^2$ parameters, we were faced 
with a dichotomy. On the one hand, our best-fitting model of $\chi_{r}^2=1.43$ had a low occurrence frequency 
(Fig.~\ref{chi2histogram}) and resulted in high eccentricities (Fig.~\ref{e-distribution}a). On the other hand, 
we encountered a significant higher occurrence frequency of larger $\chi_{r}^2$ values. At the moment we have 
no explanation for this trend. One possibility is the inability of the LM minimisation algorithm to escape 
a local minimum in the non-linear parameter space. To test this possibility, one can use a 
least-squared minimisation procedure based on a genetic algorithm (GA) or a Bayesian Markov Chain Monte 
Carlo method. 

The results of our study suggest that we should be able to trust the nominal $\chi_{r}^2=1.43$ best-fitting 
parameters as this fit represents, most 
likely, the global minimum in $\chi_{r}^2$ parameter-space, and provides the best possible description of 
the observed timing data set. However, as we have shown, the resulting best-fitting orbit with $\chi_{r}^2=1.43$ 
was unstable. Models in the neighbourhood of this system were also unstable with a collision or ejection taking place on 
very small time scales. This instability can be attributed to high values of the final fitted eccentricities for one or both 
planetary companions. It is worth mentioning that models with goodness-of-fit close to the nominal best-fitting model did not allow 
low-eccentricity/circular orbits which would have rendered them most likely to be stable. We would like to mention that 
other authors have encountered similar (in)stability problem for the two planets in the systems of NN Serpentis 
and UZ Fornacis as well \citep{Potter2011, Beuermann2010}. The two-planet system HW Virginis \citep{Lee2009} also appears to be unstable (B. Funk, private communication). In particular, it is interesting to note that the best-fitting models of 
the proposed UZ Fornax circumbinary planetary system also attained high final eccentricities \citep{Potter2011}.

Examining the $\chi_{r}^2$ frequency distribution in detail resulted in the emergence of a second population 
of models with fitted parameters that fulfill our orbital stability constraints. 
Orbital stability was examined for models with $\chi_{r}^{2}=1.89$. Although this population had a larger 
$\chi_{r}^2$ goodness-of-fit than the nominal case, we were able to find stable orbits due to significantly lower orbital 
eccentricities. Combining/pairing a least-squared minimisation algorithm with a stability analysis enabled
us to determine a new set of orbital elements for the two planets which are different from those given by 
\citet{Qian2011}. The latter is another implication of our study which points to the existence of a variety of stable 
configurations when considering models with increasingly larger $\chi_{r}^2$.

Assuming that no other physical effects are capable of explaining the observed timing variations in the light curve
of HU Aquarii, we are left with the two-planet model as the most probable cause for the observed timing variation. Our study indicates that single one-planet LTT sinusoidal timing variations are inadequate to describe the observational data. However, it is necessary to mention that timing anomalies could also be caused by direct perturbations of a circumbinary planet on the binary orbit, though these perturbations would introduce a high-frequency (short-period) component to the eclipse timing variations due to the short orbital period of the binary system. A second possibility worthy to consider in future work would be to include mutual perturbations between the two proposed circumbinary planets. Explaining eclipse timing variations caused by the latter effect has not been explored for circumbinary planetary systems. 

Finally, a similar study by \cite{Wittenmyer2011} also addresses the instability of the two proposed companions around HU Aqr from revised LTT models. In their work the authors point out the need to obtain a better understanding of the mutual interaction between the two binary components. Binary orbital period modulation caused by magnetic interaction and/or mass-transport between the two binary components could also possibly explain the observed timing anomaly for HU Aqr. However, a combination of the above effects might also provide a satisfactory description of the observed timing data. At this point, we recommend future discoveries of multi-planet circumbinary systems also include a preliminary orbital stability study as a necessary condition to increase the likelihood of the existence of such systems. At the moment it seems that best-fitting single LTT models 
in superposition ($\tau_1 + \tau_2$) are inadequate to result in stable multi-planetary circumbinary systems. Though such a model provide a good description to the observational timing data. Furthermore, we also recommend future studies to obtain and publish timing data of the secondary eclipse (if available). Any timing variations introduced from gravitational interactions and/or LTT should also provide an adequate description of timing measurements obtained from the secondary eclipse.

Further constraining of our model requires obtaining additional timing data of the HU Aqr system. Using the results of our study involving stable planetary orbits, we predict that future mid-egress timing measurements would result in $O-C$ timing differences of -10.0 to 0.0 seconds (predicted at epoch $E = 40000$ in  Fig.~\ref{BestFitOmC_2}). We emphasise that high-accuracy timing measurements are crucial to unveil the true nature of the observed timing anomalies as well as the development of improved models describing eclipse (egress) timing variations of a binary star system.

\section*{Acknowledgments}

The work by T.C.H were carried out under the Korea Astronomy and Space Science Institute Postdoctoral Research Fellowship Program. Numerical simulations were carried out on the ``Beehive'' computing cluster maintained at Armagh Observatory (UK) and the Centre for Scientific Computing at the University of Sheffield (UK) and St. Andrews University (UK). T.C.H acknowledges Martin Murphy for assistance in using the Beehive cluster. Astronomical research at Armagh Observatory is funded by the Department of Culture, Arts and Leisure (DCAL). N.H. acknowledges support from NASA Astrobiology Institute under Cooperative Agreement NNA04CC08A at the Institute for Astronomy, University of Hawaii, and NASA EXOB grant NNX09AN05G. K.G. is supported by the Polish Ministry of Science and Higher Education through grant N/N203/402739. We would like to thank Barbara Funk for sharing early results on the stability of circumbinary two-planet systems.

\bibliographystyle{mn_new}

\begin{table*}
\caption{Best-fitting parameters for the LTT orbits of HU Aqr. Subscript 
$b$ refers to orbital quantities for the compined binary. Subscript $1,2$ 
refers to the circumbinary companions. Note that the eccentricity, 
inclination and orbital period are shared quantities as outlined in the text.
The last two entries denote the minimum mass and minimum semi-major axis of the two proposed companions.}
\begin{tabular}{lcccc}
\hline
Parameter                && \multicolumn{2}{c}{two-Kepler LTT}                   &   Unit            \\ [1.5mm] \cline{3-4} \\ [-2.0ex]
                         && $\tau_{1}$                 & $\tau_{2}$            &                   \\ 

\hline
red. $\chi_{r}^2$            &&  \multicolumn{2}{c}{1.43}                          &   -               \\
\hline
$T_0$                    &&  \multicolumn{2}{c}{2,452,174.365694(2)}           &   BJD             \\
$P_0$                      &&  \multicolumn{2}{c}{0.0868204066(14)}              &   days               \\
$a_{b}\sin I_{1,2}$     &&  0.0179(5)                 &  0.0353(5)            &   au              \\
$e_b~~(\textnormal{or}~e_{1,2})$                      &&  0.075(35)                 &  0.29(3)              &   -                \\
$\omega_b$                 &&  316(2)                    &  351(1)               &   deg             \\
$T_b$                      &&  2,453,462(16)             &  2,450,050(7)         &   BJD             \\
$P_{b}~~(\textnormal{or}~P_{1,2})$                &&  2359(7)                   &  5646(8)              &   days                        \\

\hline

$K$                      && $1.03(0.03)\times 10^{-4}$ & $2.04(0.02)\times 10^{-4}$ &   days                            \\
$f(M_{1,2})$             && $1.38(0.13)\times 10^{-7}$ & $1.84(0.09)\times 10^{-7}$ &  M$_{\textnormal{Jup}}$       \\
$M_{1,2} \sin I_{1,2}$   && $5.4(0.2)$                 & $5.9(0.1)$            &  M$_{\textnormal{Jup}}$       \\
$a_{1,2}\sin I_{1,2}$          && $3.45$                     & $6.18$                &  au                       \\
\hline
\end{tabular}
\label{fitparamtable}
\end{table*}

\begin{table*}

\caption{Best-fitting parameters for the LTT orbits of HU Aqr corresponding to Fig.~\ref{BestFitOmC_2}. Subscript 
$b$ refers to orbital quantities for the compined binary. Subscript $1,2$ 
refers to the circumbinary companions. Note that the eccentricity, 
inclination and orbital period are shared quantities as outlined in the text. The last two entries denote the minimum mass and minimum semi-major axis of the two proposed companions.}
\begin{tabular}{lcccc}
\hline
Parameter                && \multicolumn{2}{c}{two-Kepler LTT}                   &   Unit            \\ [1.5mm] \cline{3-4} \\ [-2.0ex]
                         && $\tau_{1}$                 & $\tau_{2}$            &                   \\ 

\hline
red. $\chi_{r}^2$            &&  \multicolumn{2}{c}{1.89}                          &   -               \\
\hline
$T_0$                    &&  \multicolumn{2}{c}{2,452,174.3657082(5)}           &   BJD             \\
$P_0$                    &&  \multicolumn{2}{c}{0.0868204062(10)}              &   days               \\
$a_{b}\sin I_{1,2}$      &&  0.0146(7)                 &  0.0321(5)            &   au              \\
$e_b~~(\textnormal{or}~e_{1,2})$                      &&  0.11(3)                   &  0.04(3)              &   -                \\
$\omega_b$                 &&  292(2)                    &  326(1)               &   deg             \\
$T_b$                      &&  2450992(15)               &  2449837(6)           &   BJD             \\
$P_{b}~~(\textnormal{or}~P_{1,2})$                &&  2226(9)                   &  5155(8)              &   days            \\

\hline

$K$                      && $8.43(0.04)\times 10^{-5}$ & $1.85(0.01)\times 10^{-4}$ &   days                            \\
$f(M_{1,2})$             && $8.38(0.11)\times 10^{-8}$ & $1.66(0.07)\times 10^{-7}$ &  M$_{\textnormal{Jup}}$       \\
$M_{1,2} \sin I_{1,2}$   && $4.5(0.2)$                 & $5.7(0.2)$            &  M$_{\textnormal{Jup}}$       \\
$a_{1,2}\sin I_{1,2}$          && $3.32$                     & $5.81$                &  au                       \\
\hline
\end{tabular}
\label{fitparamtable_2}
\end{table*}

\end{document}